\documentstyle [a4p,epsf,12pt] {article}
\catcode`@=11 % This allows us to modify PLAIN macros.
\def \gsim{\mathrel{\mathpalette\@versim>}}
\def \lsim{\mathrel{\mathpalette\@versim<}}
\def \@versim#1#2{\lower0.4ex\vbox{\baselineskip\z@skip\lineskip\z@skip
     \lineskiplimit\z@\ialign{$\m@th#1\hfil##\hfil$%
     \crcr#2\crcr\sim\crcr}}}
\catcode`@=12 % at signs are no longer letters
\def\ee{\mathrm{e^+ e^-}}
\def\mumu{\mathrm{\mu^+ \mu^-}}

\def\pipi{\mathrm{\pi^+ \pi^-}}

\def\B0{{\mathrm{B^0}}}
\def\Bobar{\mathrm{\bar{B}^0}}

\def\DSTST{\mathrm{D^{**}}}

\def\DSPM{\mathrm{D^{*\pm}}}
\def\Ks{\mathrm{K^0_S}}
\def\Jp{\mathrm{J}/\psi}
\def\Q2j{Q_{\mathrm{2jet}}}
\def\Qop{Q_{\mathrm{opp}}^{\kappa=0.5}}
\def\Qsam{Q_{\mathrm{same}}^{\kappa=0.4}}
\def\Qvtx{Q_{\mathrm{vtx}}}
\def\sQvtx{\sigma_{Q{\mathrm{vtx}}}}
\def\Ql{Q_\ell}
\def\QB{Q_{\mathrm B}}
\def\tr{t_{\mathrm rec}}
\def\str{\sigma_t}

\def\bbbar{\mathrm{b\overline{b}}}

\def\arr{\rightarrow}

\def\dmd{\Delta m_{\mathrm{d}}}

\def\Gcs{GeV/$c^{2}$}

\def\Gc{\rm GeV/$c$}
\def\etal{{\sl et al.}}
\def\ec2{E_{\mathrm cone}}
\def\eco2{E_{\mathrm cone2}}
\newcommand{\stwob}{\mbox{$\sin 2\beta$}}
\newcommand{\acp}{\mbox{$\sin 2\beta$}}

\newcommand{\epostfig}[3]{
\begin{figure}[tbp]
\setlength{\epsfxsize}{1.1\hsize}
\hspace*{-0.05\hsize} \epsfbox{#1}
\caption{\label{#2}#3}
\end{figure}
}
\newcommand{\PRL}[3] {Phys.~Rev.\ {Lett.~{\bf #1}} (#2) #3}
\newcommand{\PRD}[3] {Phys.~Rev.\ {\bf D #1} (#2) #3}

\newcommand{\ZPC}[3] {Z.~Phys.\ {\bf C #1} (#2) #3}

\newcommand{\PLB}[3] {Phys.~Lett.\ {\bf B #1} (#2) #3}
\begin{document}           % End of preamble and beginning of text.
%================================================== TITLE =======
\begin{titlepage}
\begin{center}
  {\large   EUROPEAN LABORATORY FOR PARTICLE PHYSICS }
\end{center}
\bigskip
\begin{tabbing}
\` CERN-EP/98-001 \\
\` 16th January 1998 \\
\end{tabbing}
%======================================================== TITLE =======
\vspace{1 mm}
\begin{center}{\LARGE\bf
Investigation of CP violation in 
\mbox{\boldmath $\B0 \arr \Jp \Ks$} decays
at LEP
} \end{center}
\vspace{10mm}
\begin{center}{\LARGE
The OPAL Collaboration
}\end{center}
%\begin{center} {\bf Authors : Elisabetta Barberio, Richard Hawkings
%    and Martin Jimack}
%\end{center}
%\begin{center} {\bf Editorial board : Richard Batley, 
% Peter Igo-Kemenes, Hassan Jawahery, Giora Mikenberg}
%\end{center}
\vspace{6 mm}
%==============================================  start   abstract =====
\begin{abstract}
\vspace{5 mm}
An investigation of CP violation was performed
using a total of 24 candidates for $\B0 \arr \Jp \Ks$ decay, 
with a purity of about 60\%.
These events were
selected from 4.4 million hadronic $\rm Z^0$ decays 
recorded by the OPAL detector at LEP.
An analysis procedure, involving
techniques to reconstruct the proper decay times
and tag the produced b-flavours, $\B0$ or $\Bobar$,
has been developed 
to allow a first direct study of the time dependent CP asymmetry
that, in the Standard Model, is $\sin 2\beta$.
The result is
\[
\sin 2\beta  = 3.2 _{-2.0}^{+1.8}\ \pm 0.5 \; ,
\]
where the first error is statistical and the second systematic.
This result is used to determine probabilities for different
values of $\sin 2\beta$ in the physical region from $-1$ to +1.
\end{abstract}
%================================================ end of abstract
%=====
\vspace{10mm}
\vspace{10mm}
\vspace{30mm}
\begin{center}
(Submitted to Physics Letters B)
\end{center}
\end{titlepage}
\begin{center}{\Large        The OPAL Collaboration
}\end{center}\bigskip
\begin{center}{
%begin authorlist
K.\thinspace Ackerstaff$^{  8}$,
G.\thinspace Alexander$^{ 23}$,
J.\thinspace Allison$^{ 16}$,
N.\thinspace Altekamp$^{  5}$,
K.J.\thinspace Anderson$^{  9}$,
S.\thinspace Anderson$^{ 12}$,
S.\thinspace Arcelli$^{  2}$,
S.\thinspace Asai$^{ 24}$,
S.F.\thinspace Ashby$^{  1}$,
D.\thinspace Axen$^{ 29}$
% ...
%end authorlist
}\end{center}\bigskip
\bigskip
%begin institutes
$^{  1}$School of Physics and Astronomy, University of Birmingham,
Birmingham B15 2TT, UK
\newline
$^{  2}$Dipartimento di Fisica dell' Universit\`a di Bologna and INFN,
I-40126 Bologna, Italy
\newline
% ...
$^{  5}$Cavendish Laboratory, Cambridge CB3 0HE, UK
\newline
% ...
$^{  8}$CERN, European Organisation for Particle Physics,
CH-1211 Geneva 23, Switzerland
\newline
$^{  9}$Enrico Fermi Institute and Department of Physics,
University of Chicago, Chicago IL 60637, USA
\newline
% ...
$^{ 12}$Indiana University, Department of Physics,
Swain Hall West 117, Bloomington IN 47405, USA
\newline
% ...
$^{ 16}$Department of Physics, Schuster Laboratory, The University,
Manchester M13 9PL, UK
\newline
% ...
$^{ 23}$Department of Physics and Astronomy, Tel Aviv University,
Tel Aviv 69978, Israel
\newline
$^{ 24}$International Centre for Elementary Particle Physics and
Department of Physics, University of Tokyo, Tokyo 113, and
Kobe University, Kobe 657, Japan
\newline
% ...
$^{ 29}$University of British Columbia, Department of Physics,
Vancouver BC V6T 1Z1, Canada
\newline
% ...
\newpage
%end notes
%%%%%%%%%%%%%%%%%%%%%%%%%%%%%%%%%%%%%%%%%%%
\section[intro]{Introduction}
CP violation was observed in 1964 in $\mathrm K^0$ decays \cite{cronin},
and, so far, this phenomenon
has been seen only in the K system.
CP violation can be accommodated
in the Standard Model, provided that the CKM matrix elements
are allowed to be complex.
The CP violating effects associated with b hadrons
are expected to be 
larger than in the K system~\cite{cpgold}.
It is therefore important to investigate   
CP violation in the B system, 
where the relation between CKM matrix elements and CP violation
can be tested.
%since this provides powerful
%constraints on the CKM matrix elements.
Previous studies, yielding null results~\cite{cpb,inclept},
have focussed on CP violation in inclusive B decays, 
predicted to be at the level of $10^{-3}$.

This paper presents a study of CP violation in
$\B0 \arr \Jp \Ks$ decays.
The decay mode $\Jp \Ks$ has long been considered
a `golden' channel for CP violation studies \cite{cpgold,bigi},
since the final state is a CP eigenstate
which is experimentally favourable for reconstruction
because the $\Jp$ and $\Ks$ are narrow states
and $\Jp \arr \ell^+ \ell^-$ decays give a distinctive 
signature.
In addition, CP violation in this channel is dominated by
diagrams having a single relative phase, allowing a clean
extraction of the phase of a CKM matrix element.
In the Standard Model,
the expected time-dependent rate asymmetry, $A(t)$, 
is given by 
\begin{equation} 
A(t) \equiv \frac{\B0 (t) - \Bobar (t)}{\B0 (t) + \Bobar (t)} 
= -\stwob \sin \dmd t \; , 
\end{equation} 
where the parameter $\dmd$ is the mass difference between the two
$\B0$ mass eigenstates and
%\begin{equation} 
%A(t) \equiv \frac{\B0 (t) - \bar{\B0} (t)}{\B0 (t) + \bar{\B0} (t)}
%\end{equation}
$\B0 (t)$ ($\Bobar (t)$) represents the
rate of produced $\B0$'s ( $\Bobar$'s ) 
% as a function of the 
decaying to $\Jp \Ks$ at a given
proper decay time, $t$. 
%In the Standard Model, $\asim=-\stwob$, where 
The angle $\beta$ is given by
\begin{equation}
\beta \equiv \arg \left [ \frac{V_{\rm td} V_{\rm tb}^*}{V_{\rm cd}
    V_{\rm cb}^*} \right ]
\approx -\arg V_{\rm td} \; .
\end{equation}
Constraints on the CKM matrix, including
measurements of CP violation in the K system,
imply that the Standard Model expectation for
$\sin 2\beta$ lies 
in the range 0.3--0.9~\cite{cpgold}.
Other models of CP violation, such as the Superweak
model~\cite{sweak}, would give a time dependence of the same
form, but with $\sin 2\beta$ replaced by another amplitude
of magnitude less than or equal to one.
%with the magnitude of \asim\ less than or equal to one.

At LEP, due to the
small branching ratios of $\B0 \arr \Jp \Ks$ and $\Jp \arr \ell^+ \ell^-$,
only a handful of these decays may be seen.
It is therefore important to maximise the reconstruction efficiency 
and
to determine the b-flavour at production with a minimum error rate.
In contrast, the proper time resolution is not critical, since 
the frequency of $\B0$ oscillation is easily resolved.
In this analysis, $\B0\arr\Jp\Ks$ decays are reconstructed and their
decay proper times are measured. 
The production flavour ($\B0$ or $\Bobar$) 
of each candidate is determined using a combination of jet and
vertex charge techniques. 
The CP-violating amplitude, $\sin 2\beta$, is extracted using an unbinned
maximum likelihood fit to the proper-time distribution of the
selected data. Such a fit has higher expected sensitivity than a
time-integrated fit, even if the domain of integration of the latter is
optimised, since it uses the additional, well measured, decay-time
information.

The next section describes the event selection and proper time
estimation.
Section 3 describes the tagging of the production flavour.
In section 4, the fits and results are presented,
with systematic errors discussed in section 5.
Discussion of the result and conclusions are given in section 6.

%\section[select]{Event selection}
\section[select]{\mbox{\boldmath $\B0 \arr \Jp \Ks$} reconstruction}

A detailed description of the OPAL detector may be found elsewhere
\cite{opaldet}.
The basic selection of $\B0 \arr \Jp \Ks$ decays
was described in a previous publication~\cite{rison},
which used data collected between 1990 and 1994
to identify various exclusive B decay modes.
%To summarise, $\Jp$ particles were identified by 
%their decays to $\mumu$ or $\ee$, where the 
%invariant mass of the lepton pair was required
%to be in the range 2.9--3.3\,\Gcs .
%The $\Ks$ candidates were identified through their
%decays to $\pipi$ in a procedure based on a search
%for a displaced $\mathrm V^0$ vertex.
%The $\pipi$ invariant mass was required to lie
%in the range 0.467--0.527\,\Gcs .
%The $\B0$ candidate was reconstructed using 
%the SQUAW package~\cite{squaw} to
%constrain the track quantities in a kinematic
%fit.
%A total of 10 $\B0 \arr \Jp \Ks$ candidates were
%selected with masses in the range 5.15--5.40\,\Gcs ,
%with an estimated background of 1.6 events.
For this analysis, the 1995 data were included
to give a total of 4.4 million events passing the basic
hadronic event selection.
The efficiency of the $\B0 \arr \Jp \Ks$ selection was also improved
(and applied to the full data sample) by relaxing or modifying the criteria, 
at the expense of a somewhat larger background.
The efficiency and purity of the selection was studied using
Monte Carlo events generated with Jetset 7.4~\cite{jetset}
and processed through the OPAL detector simulation~\cite{gopal}.

Lepton candidates were required to satisfy 
the polar angle\footnote{
The right-handed coordinate system is defined with
positive $z$ along the $\mathrm{e}^-$ 
beam direction, $x$ pointing to the centre of the LEP ring, $\theta$ and
$\phi$ as the polar and azimuthal angles,
and $r^2 = x^2 + y^2$.
The origin is taken to
be the centre of the detector.}
cut $|\cos\theta | < 0.97$
and to have track momenta larger than 1.5~\Gc\ (2~\Gc\ )  for muon
(electron) candidates.
% 
%For the $\Jp \arr \mumu$ selection,
%the muon identification was relaxed.
Muons were identified by requiring
%The positional matching requirement between 
an extrapolated track to match the position of
a muon chamber segment %was loosened from 1.5 to 
to within 4 standard deviations.
Muon candidates were also considered if
the muon chamber segment had no $z$-coordinate
reconstructed,
provided that the matching in position and angle were
within 4 standard deviations in the $r$-$\phi$ plane. 
As in \cite{rison}, muons identified in the hadronic 
calorimeter were accepted in regions without muon chamber
coverage.

%In $\Jp \arr \ee$ decays, 
Electrons were identified~\cite{enet}
and photon conversions rejected~\cite{nnconv} 
using artificial neural network algorithms.
%The output of the network used for the
%electron selection was required to be larger
%than 0.8.
%
When the electron energies are determined only from
the reconstructed track momenta,  
photon radiation causes the reconstructed $\Jp$ mass
spectrum to have a long tail to lower masses, reducing the
efficiency of a mass cut.
Therefore the electron energies were determined using in addition the
information from the lead-glass calorimeter.
The energy contained in a cone of 30\,mrad around
the impact point of the electron track on the calorimeter,
%the projection of the electron track into the calorimeter,
plus energy contained in lead-glass blocks touching this
cone, were summed~\cite{elid}.
This sum was used as the  energy of the electron
if larger than the track momentum, otherwise the track 
momentum was used.
%In addition, 
%Conversions were rejected using a neural network algorithm~\cite{nnconv},
%which is more efficient than the algorithm used before~\cite{rison}.

$\Jp$ candidates were selected by demanding two electron or two muon
candidates of 
opposite charge with an opening angle of less than $90^\circ$.
The invariant mass of the two leptons was required to lie in the range
2.95--3.25\,\Gcs\ for $\Jp\arr\mumu$ candidates and 2.95--3.40\,\Gcs\
for $\Jp\arr\ee$ candidates
(in the latter case over-correction for 
%the photon radiation correction causes
photon radiation causes
a significant tail at higher masses).
% in the electron channel. 
%{\bf do we need a plot of the mass spectrum 
%in data to justify all the new cuts?}.

The $\Ks$ selection was based on the procedure described
previously~\cite{Ks}, considering the intersection of all track
pairs of opposite charge (excepting $\Jp$ candidate tracks)
passing certain quality criteria.
%As in~\cite{rison}, no requirements were made on the 
%impact parameters
% of the tracks with respect to the  beam spot position.
The projection of the
the $\Ks$ momentum vector in the $r$-$\phi$ plane 
was required to point back to the beam spot position
to within $8^\circ$.
The beam spot position was measured using charged
tracks from many consecutive events,
thus following any significant shifts in beam position
during a LEP fill~\cite{beamspot}. 
%this cut was relaxed
%to $8^\circ$, and  
The impact parameter\footnote{
The impact parameter of a track with
respect to a vertex
is defined as
the distance of closest approach of the track
to that vertex.} significance 
(the impact parameter divided by its error)
%the distance of
%closest approach of the $\Ks$ momentum vector to the $\Jp$
%candidate vertex, 
in the $r$-$\phi$ plane of the $\Ks$
with respect to the $\Jp$ vertex
%divided by its error, 
was required to be less than~5.
The reconstructed distance between the $\Jp$ vertex and the $\Ks$
decay vertex, divided by its error, was required to exceed~2. 
%This
%replaces a cut requiring the $\Ks$ decay vertex to be more than 2\,cm
%from the nominal interaction point.
If the $\Ks$ decay vertex was inside the active volume of the jet
chamber ($r>30$\,cm), the radial coordinate of the first jet chamber hit on
either track was required to be at most 10\,cm less than the decay
vertex radial coordinate, 
and the tracks were required not to have any associated
vertex chamber or silicon microvertex detector hits.
The invariant mass of the $\Ks$ candidate was required to lie in the
range 0.45--0.55\,\Gcs . 
%rather than 0.467--0.527\,\Gcs\ as previously. 
In order to suppress a potential background from 
$\Lambda_{\rm b}\arr\Jp\Lambda$ decays, the $\Ks$ candidate was rejected if
its invariant mass under either proton-pion hypothesis was in the range
1.110--1.121\,\Gcs .

$\B0$ candidates were reconstructed by combining $\Jp$ and $\Ks$
candidates from the same thrust hemisphere\footnote{
The two hemispheres were separated by the plane perpendicular to the
thrust axis of the event and containing the $\rm e^+e^-$ interaction point.}.
Kinematic fitting using the SQUAW
package~\cite{squaw}, constraining the $\Jp$ and $\Ks$
masses to their nominal values, was employed, %as before, 
and the
probability of the kinematic fit was required to exceed 1\%.
The invariant masses of selected $\B0$ candidates were required to lie in
the range 5.0--5.6\,\Gcs ,
and the $\B0$ energies %was required 
to exceed 20\,GeV. 
The efficiency for reconstructing the decay 
$\B0\arr\Jp\Ks\arr\ell^+\ell^-\pipi$
was estimated to be $17.3\pm 1.4$\% where the error is due to Monte Carlo 
statistics. 
This compares to 10.9\% for the previous selection~\cite{rison}.

The distribution of reconstructed mass  
is shown in Figure~\ref{f:mass}(a) for
candidates passing the entire selection
except the $\Jp \Ks$ mass cut.
In total, 24 candidates were selected in the mass region 5.0--5.6\,\Gcs.
The background was estimated from the data using an
unbinned maximum likelihood
fit to the joint distribution of the
$\Jp \Ks$ mass and energy, including mass values  
between 4 and 7\,\Gcs .
The shape of the mass distribution 
for the signal was taken from Monte Carlo,
and parametrised using
two Gaussians for the peak of the distribution and a third to account
for a significant tail to lower masses.
The shape of the background mass distribution was
also taken from Monte Carlo, and parametrised using a polynomial function.

%The signal is represented as a Gaussian of width 
%50\,\Mcs (corresponding to the mass resolution seen in Monte Carlo),

%The background was parametrised by a negative exponential.
%Reflections (where one
%particle is misidentified as another) and satellites (where one or more
%5particles are missed) could cause the background to peak
%in the region below the B mass, rather than smoothly diminishing with
%mass,
%and would therefore cause an error in the estimated
%background level in the signal region.
%Such effects are treated as systematic uncertainties. 
%The possible effects of reflections (where one
%particle is misidentified as another) and satellites (where one or more
%particles are missed) are treated as systematic uncertainties.
%Monte Carlo
%studies showing no significant peaks from reflections (where one
%particle is misidentified as another) or satellites (where one or more
%particles are missed) in the mass range being fitted. 
%
\epostfig{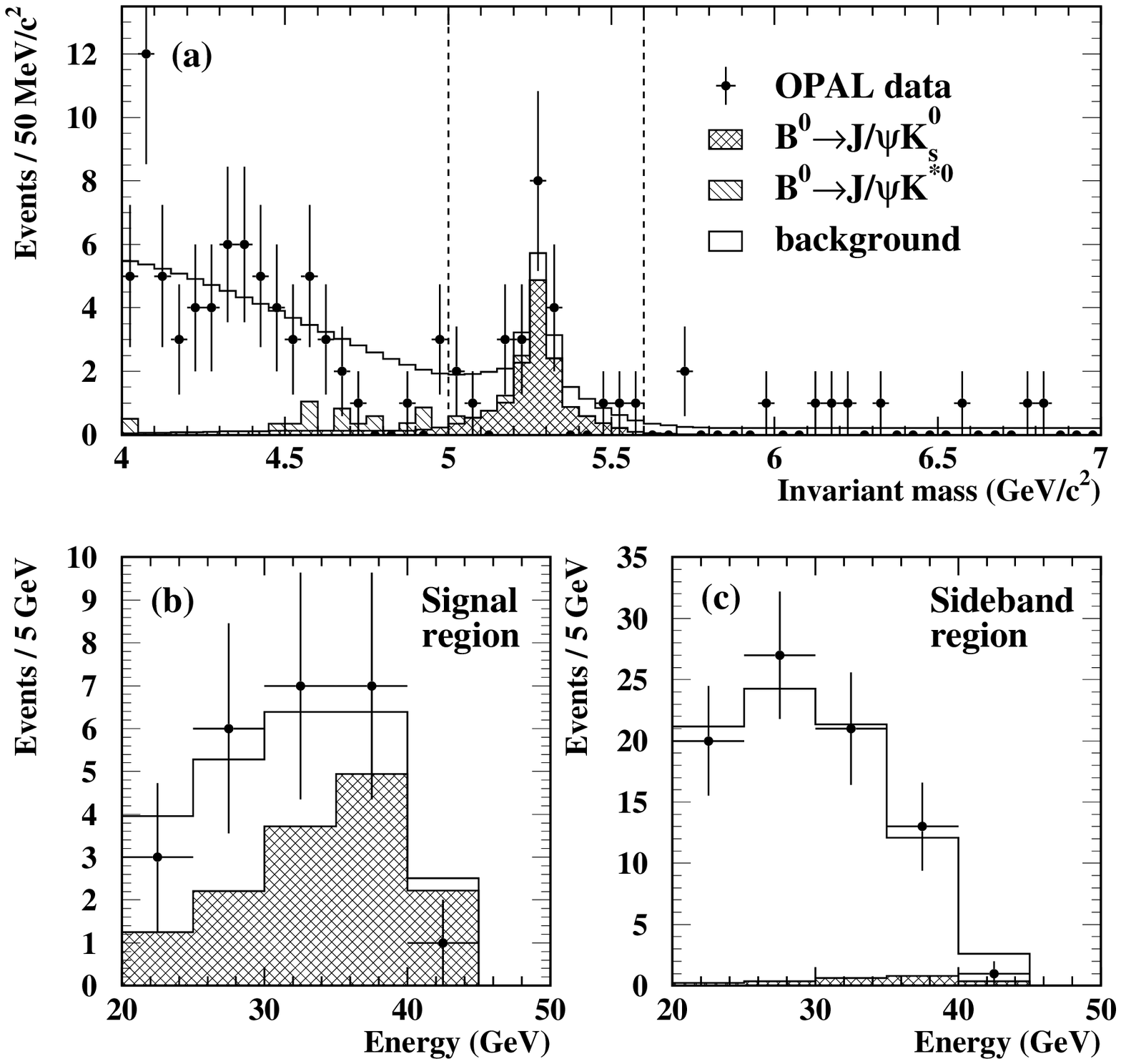}{f:mass}{
(a) The mass distribution of $\Jp \Ks$ candidates.
(b) The energy distribution of the candidates with masses between
5.0 and 5.6\,\Gcs.
(c) The energy distribution of the candidates with masses outside
this region.
In each case,
the data are shown by the points with error bars, and the 
fit is shown by the open histogram. 
The estimated contribution from the
$\B0\arr\Jp\Ks$ signal %being hatched.}
is shown by the cross hatched histogram. The estimated contribution
from $\B0\arr\Jp K^{*0}$ is also shown in (a) by the hatched
histogram.
Note that the $\Jp K^{*0}$ contribution is taken directly from 
Monte Carlo, while the shapes of the other distributions
are parametrised (see text).
}

%The energy distributions of both signal and background were
%taken to be uncorrelated with the mass distributions, 
%which is consistent
%with the Monte Carlo.
%The energy distribution of the signal was taken from Monte Carlo,
%parametrised using a Peterson fragmentation function~\cite{Peterson}.
%The background was also parametrised using a Peterson function,
%but with the shape parameter, $\epsilon$, allowed to vary.
%
The mass and energy distributions were taken to be uncorrelated both
for the signal and the background, 
since the correlations seen in Monte Carlo
were small and had a negligible effect on the fit.
Peterson fragmentation functions~\cite{Peterson} were used to
parametrise the energy distributions. For the signal, the
Peterson parameter
$\epsilon$ was taken from a fit to the Monte Carlo energy
distribution, while for the background it was allowed to vary. % freely.
In total, three parameters
were allowed to vary in the fit: the number of signal candidates, the
position of the $\B0$ mass peak
and the Peterson parameter for the background energy
distribution.
The result of the fit is shown in Figure~\ref{f:mass}(a) as a function
of mass, 
in Figure~\ref{f:mass}(b) as a function of energy
for masses between 5.0 and 5.6\,\Gcs\ (signal region)
and in Figure~\ref{f:mass}(c) as a function of energy
for masses outside this region (sideband).
The overall fitted purity of the 24 $\B0\arr\Jp\Ks$ candidates is 
$60 \pm 8 \%$, where the error is statistical.
%due to the data statistics.
The fit was used not only to determine the overall
purity of the sample, but also to assign event-by-event 
background probabilities, $f_{\mathrm bac}$,
to be used in the fit for \acp ,
according to the reconstructed $\Jp\Ks$ mass and energy of each candidate.

The $\B0$ decay length, $l_\B0$, was determined from the distance
between the average beam spot position and the $\Jp$ vertex in the
$x$-$y$ projection, 
%constrained to lie along the projection of 
%using 
%converted into 3 dimensions 
correcting for the polar angle
using the direction of the $\Jp\Ks$
momentum vector.
%as a constraint.  
The $\B0$ momentum, $p_\B0$, was taken from the constrained fit of the
$\Jp \Ks$ system, and the proper decay time was then calculated as
\begin{equation} 
\tr = l_\B0 \cdot \frac{M_\B0}{p_\B0} \; ,
\end{equation}
where $M_\B0$ is the $\B0$ mass, taken to be 
5.279\,\Gcs\ ~\cite{pdg96}.
The uncertainty on $\tr$, 
typically $\str=0.1$\,ps,
was estimated by combining the uncertainties on $l_\B0$
and $p_\B0$. 
%and was typically 0.1\,ps. 
%The mean decay proper time of the 16
%candidate events was found to be $1.4\pm 0.3$\,ps, consistent with
%the world average $\B0$ lifetime of $1.56\pm 0.06$\,ps \cite{pdg96}.

\section[tag]{Tagging the produced b-flavour}

Information from the rest of the event was used to determine the
production flavour of each candidate. 
The weighted
track charge sum, or `jet charge', gives information on the charge,
and hence b-flavour, of
the primary quark initiating the jet 
within which the $\Jp \Ks$ candidate was isolated.
Additionally, since the $\rm Z^0$ decays into
quark-antiquark pairs, measuring the b-flavour of the other b quark
produced in the event also provides information on the production
flavour.

In this analysis, three different pieces of information were used 
to determine the $\B0$ production flavour\footnote{
Leptons in the opposite hemisphere were also investigated
as a possible tag, but in this sample the
events with selected leptons were found to have large %values of
jet charges (in agreement with the lepton charge), 
so that there was no significant gain in 
%also using leptons to tag
% the production flavour 
tag purity from the use of leptons.}:
(a) the jet charge of the highest energy jet other
than that containing the $\B0$ candidate,
(b) the jet charge of the jet containing the $\B0$ candidate,
 excluding the tracks from the $\Jp$ and $\Ks$ decays, and
(c) the vertex charge of a significantly separated vertex (if existing)
in the opposite hemisphere.

Jets were reconstructed from tracks and 
electromagnetic clusters not associated to tracks using the JADE E0
recombination scheme~\cite{jetfinder} with a $y_{\rm cut}$ value of 0.04.

The jet charge for the highest energy jet other than the one
containing the $\Jp \Ks$ candidate was calculated  as :
%by summing the track charge, weighted by
%the longitudinal component of the track momentum with respect to the
%jet axis $p^l_i$ divided by the beam energy $E_{\rm beam}$, 
%raised to the power $\kappa$:
\begin{equation} 
Q_{\mathrm{opp}}=\sum_i \left( \frac{p^l_i}{E_{\rm beam}}\right) ^\kappa
q_i \; ,
\end{equation}
where $p^l_i$ is the longitudinal component of the  momentum of track
$i$ with respect to the jet axis, $E_{\rm beam}$ is the beam energy, $q_i$ 
is the electric charge ($\pm 1$) of each track
and the sum is taken over all tracks in the jet.
Using simulated data, the 
value of $\kappa$ that minimised the mistag probability (the
probability to tag a $\B0$ as $\Bobar$ and vice versa), 
was found to be approximately 0.5.
The mistag probability includes the effect of $\B0$ mixing in this
jet.
The jet charge for the jet containing the $\B0$ candidate, $Q_{\mathrm{same}}$,
was calculated in the same way, but excluding
the $\Jp$ and $\Ks$ decay products.
These particles 
contain no information on
whether their parent was produced as a $\B0$ or $\Bobar$,
and would only dilute the information from the fragmentation tracks.
The optimal value of $\kappa$ was found to be 0.4 in this case,
smaller than that for the other b-hadron as the high momentum B decay
products were excluded.

Secondary vertices were reconstructed in jets in the 
hemisphere opposite to the $\B0$ candidate 
in data where silicon microvertex information was available,
using the algorithm described 
in \cite{opalrb}. 
For the 1991 and 1992 data, vertices were
reconstructed in the $x$-$y$ plane only. 
In the 1993--1995 data, precise $z$ coordinate information from the
silicon microvertex detector was also available, and vertices were
reconstructed in three dimensions using an extension of the vertex
finding algorithm as in \cite{higgs}. 
%No attempt was made to reconstruct vertices %in data
%without silicon information 
%in the hemisphere opposite to the $\B0$ candidate.
A secondary vertex was accepted if the distance from the primary to
the secondary vertex divided by its error (the vertex significance) exceeded 3.
If more than one vertex in the opposite hemisphere satisfied 
this requirement %$L/\sigma_L>3$,
the one with the highest significance
%value of $L/\sigma_L$ 
was taken.
Approximately 40\% of Monte Carlo $\B0\arr\Jp\Ks$ events had 
such an accepted %selected 
secondary vertex in the opposite hemisphere.
For the selected vertex,
%For these vertices, 
the charge $\Qvtx$ and its uncertainty
$\sQvtx$ were calculated using an improved version of the algorithm
described in~\cite{bobK}. 
%The weight $w_i$ 
For each track $i$ in the jet 
containing the vertex,
%to belong to the secondary vertex was calculated using 
the track momentum,
the momentum transverse to the jet axis, and the track impact
parameters with respect to the primary and secondary vertices in the
$r$-$\phi$ and $r$-$z$ planes (the $r$-$z$ information was only used in
1993--1995 data)
%The information was 
were
combined to form a weight $w_i$
using an artificial
neural network algorithm. 
The weight $w_i$ quantifies the probability for track $i$
to belong to the selected secondary vertex.
The vertex charge is then calculated as:
\begin{equation} 
\Qvtx=\sum_i w_iq_i \; ,
\end{equation}
% where $q_i$ is the charge ($\pm 1$) of each track in the jet. 
and the 
uncertainty 
%was similarly calculated 
as:
\begin{equation} 
\sQvtx^2=\sum_i w_i(1-w_i)q_i^2 \; .
\end{equation}

For events with such a selected secondary vertex
(9 of the 24 $\B0$ candidates fall into this category),
a neural network was constructed to %separate 
tag the produced $\B0$
or $\bar{\B0}$, 
combining the four inputs, 
$\Qvtx$, $\sQvtx$, $\Qop$ and $\Qsam$.
The network was trained on a large Monte Carlo sample,
and a variable 
\begin{equation}\label{qbdef}
\QB(x) = {f_{\B0}(x)-f_{\Bobar}(x)} % {f_{\B0}(x)+f_{\Bobar}(x)} 
\end{equation}
was calculated as a function of the network output, $x$, 
where $f_{\B0}(x)$ ($f_{\Bobar}(x)$)
is the fraction of candidates at a particular value of $x$ 
due to produced $\B0$ (${\Bobar}$) according to Monte Carlo
(which included the effects of $\B0$ mixing).
The variable $\QB$ represents the effective produced b-flavour 
for each candidate
($\QB= +1$ or $-1$ for pure $\B0$ or $\Bobar$, respectively),
and $|\QB| = 1 - 2\eta$ 
%is equal to 
the tagging dilution, %$D = 1 - 2\eta$,
where $\eta$ is the probability to tag the production flavour
incorrectly.
The average value of $|\QB|$ is 0.38 for such events,
according to Monte Carlo.
%The value of $\QB$ represents the
%mistag or `tagging dilution' of each event---values close to +1 (-1)
%represent events very
%well tagged as $\B0$ ($\Bobar$), whilst values close to zero indicate 
%events which are very poorly tagged and equally likely to be
%$\B0$ or $\Bobar$.
%The distribution of $\QB$ 
%(after removing any asymmetry in the charge variables)
%is shown for $\B0$ and ${\Bobar}$ secondary vertex
%tagged events in Figure~\ref{f:mist}(a).
%Nine of the 24 data candidate events have a selected secondary
%vertex in the opposite hemisphere. 

For events without such a selected secondary vertex, 
only the jet charge information is available. 
In
this case the two jet charges were combined linearly to form
\begin{equation}
 \Q2j = \Qsam - 1.43\cdot \Qop \; , 
\end{equation}
where the factor of 1.43 was optimised using simulated data to
minimise $\eta$.
%overall production flavour mistag in these events. 
Distributions $f_{\B0}(\Q2j)$ and $f_{\Bobar}(\Q2j)$ were
formed and $\QB(\Q2j)$ was determined by analogy to 
equation~\ref{qbdef}.
For these events, the Monte Carlo predicts the
average value of $|\QB|$ to be 0.31.
%The distribution of $\QB$ from the $\Q2j$ distribution
%is shown for $\B0$ and $\bar{\B0}$ in
%Figure~\ref{f:mist}(b) (after removing any charge asymmetry).
%
%\epostfig{misstag.eps}{f:mist}{The distribution of $\QB$ for $\B0$
%  (solid line) and $\Bobar$ (dotted line) in Monte Carlo, 
%  from the neural network for
%  events with a selected secondary vertex (a), and from $\Q2j$ for
%  events without a selected secondary vertex (b).}

It is important to ensure that the tagging dilution
arising from
the jet and vertex charges,
which are used to determine the $\QB$ values,
is correctly modelled by the Monte Carlo.
%Monte Carlo correctly models the
%quantities used in the production flavour tagging, as the resulting 
%$f_{\B0}$ and $f_{\Bobar}$ distributions are used to calculate 
%the tagging dilution 
%$\QB$ values for each candidate in the data. 
%An error in estimating the tagging
%dilution can lead to an under or overestimate of the size and
%significance of any CP asymmetry seen in the data.
%
% The scale of the $\QB$ distributions (i.e. the tagging dilution)
The tagging dilutions arising from $\Qop$ and 
from $\Qvtx$ were checked using
large samples of data and Monte Carlo inclusive lepton events,
selected as in \cite{inclept}.
In such events, the charge of the
lepton (usually from a semileptonic decay of a b hadron), 
%and the lepton charge 
is strongly
correlated with the produced b-flavour. 
%in the opposite hemisphere. 
The distributions 
of $\Qop$ and $\Qvtx$, multiplied by the lepton charge $\Ql$, 
%for the
%product of the lepton charge with $\Qop$ and with $\Qvtx$
were compared for data and Monte Carlo,
and found to be consistent (see Figures~\ref{f:qbdist}(a) and (b)).
The mean values of these distributions 
(which are sensitive to the tagging dilution of $\Qop$ or $\Qvtx$) 
for data 
divided by those in the Monte Carlo, were found to be
$1.13 \pm 0.09$ and $1.05 \pm 0.11$ for $\Qop$ and $\Qvtx$
respectively. 
%These %observed level of agreement is 
%ratios are 
%used to evaluate
%the systematic errors from the tagging dilution.
The $\Qsam$ dilution was checked using samples
of $\B0 \arr \DSPM \ell$ candidates selected in data and Monte Carlo, 
as in \cite{charlton}.
(In calculating $\Qsam$, tracks from the $\DSPM \ell$ combinations
were excluded.)
%The $\B0$ decay products were excluded from the calculation,
%as for the $\Jp \Ks$ candidates.
%The dilution was found to be in agreement with the Monte Carlo
%prediction, with a relative statistical error of 20\%. 
The average dilution seen in the data divided by that observed in Monte
Carlo was found to be $0.98 \pm 0.20$.
The
distributions of $\Ql\cdot\Qsam$ for data and Monte Carlo are shown in
Figure~\ref{f:qbdist}(c). Only events with reconstructed $\B0$ decay
time $\tr <2\,$ps are included, to reduce the tagging dilution due to 
$\B0$ mixing.

\epostfig{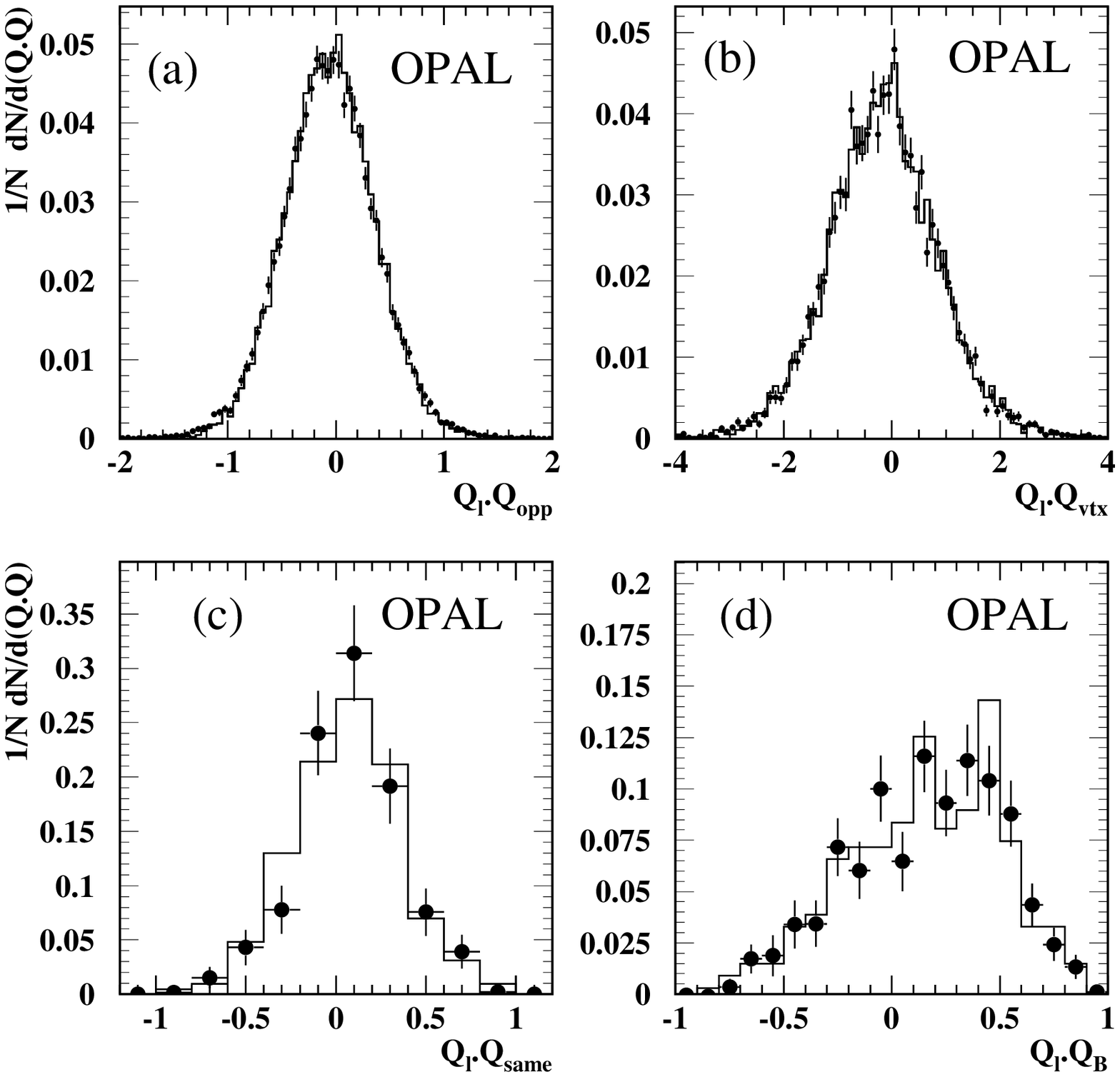}{f:qbdist}
{The distribution of the product of the lepton charge $\Ql$ and
(a) the $\Qop$ jet charge, (b) the $\Qvtx$ vertex charge,
(c) the $\Qsam$ jet charge and (d) the $\QB$ tagging
variable. Distributions (a) and (b) are taken from an inclusive lepton
sample, whilst distributions (c) and (d) are taken from a
$\B0\arr\DSPM\ell$ sample, only including events with reconstructed
proper time $\tr <2$\,ps. The data are shown as points and the Monte
Carlo predictions as histograms.}
%
%The tagging dilution of the same side jet charge $\Qsam$ is more
%difficult to test, since the $\B0$ decay products must be excluded
%from the jet charge calculation, as for the $\Jp\Ks$ candidates.
%The $\Qsam$ dilution was checked using a sample
%of $\B0 \arr \DSPM \ell$ events selected in the data, 
%as in \cite{charlton}, where the $\B0$ decay products are 
%reconstructed and excluded from the jet charge.
%The dilution was found to be in agreement with the Monte Carlo
%prediction, with a relative statistical error of 20\%.

%As a cross check,
The samples of 
$\DSPM \ell$ events were also used to check 
%the modelling of the 
the tagging dilution
of the $\QB$ values 
(formed from the combinations of jet and vertex
charges). 
The average dilution seen in the data divided by that seen in Monte
Carlo was found to be $0.96 \pm 0.14$. The distributions of
$\Ql\cdot\QB$ are shown in Figure~\ref{f:qbdist}(d), again for events
with $\tr <2\,$ps.
Both Figures~\ref{f:qbdist}(c) and ~\ref{f:qbdist}(d) show
agreement between data and Monte Carlo.

%values measured in data and Monte Carlo were found
%to be in agreement, indicating that the performance of the $\QB$
%dilution estimater is well modelled in the Monte Carlo.

The $\Qvtx$, $\Qop$ and $\Qsam$ distributions are not 
necessarily charge symmetric
because of detector effects causing
a difference in the rate and reconstruction of
%an asymmetry between 
positive and negative tracks.
These effects are caused by the material in the detector
and the Lorentz angle in the jet chamber.
They were removed by subtracting offsets from the $\Qvtx$,
$\Qop$ and $\Qsam$ values before the $\QB$ tagging dilutions were
calculated.
The $\Qvtx$ and $\Qop$ offsets were determined using
%a large sample of inclusive lepton events selected from data.
the inclusive lepton events selected from data.
The $\Qsam$ offset was determined from Monte Carlo $\B0$ jets,
since no pure sample of fully reconstructed $\B0$
decays is available from the data.
However, the $\DSPM \ell$ events do allow this offset to be checked,
even though some extra tracks may be present from $\DSTST$ decays.
The data and Monte Carlo were found to be in good agreement.
%The $\Qsam$ offset is a source of systematic uncertainty. 
%
The normalised offsets (the offsets divided by the r.m.s.~widths)
of the charge distributions
were found to be $+0.029 \pm 0.011$, $+0.018 \pm 0.007$ 
and $+0.036 \pm 0.018$
for $\Qvtx$, $\Qop$ and $\Qsam$ respectively. 
The error
quoted for the $\Qsam$ offset is the statistical precision
of the $\DSPM \ell$ events in data.
If these offsets were not removed, they
would induce respective shifts of $-0.004$, $-0.003$ and $+0.009$
in the overall $\QB$ distribution.
%If all the offsets were removed simultaneously,
If no corrections were applied for the offsets,
the combined shift would be $+0.002$.
%is a few percent of the 
%width of the distribution for all three of the charge variables. 
%and statistical precision 
%to which it is  known gives rise to systematic error.

\section[results]{Fit and results}

The reconstructed proper time $\tr$ and tagging variable $\QB$ of each
of the 24 candidates is shown in Figure~\ref{f:fit}(a).
The events with invariant mass in the range
5.15--5.40\,\Gcs\  are shown as filled circles, whilst the events
in the ranges 5.00--5.15 and 5.40--5.60\,\Gcs, which have lower 
signal purity, are shown as open circles.

In order to quantify the CP asymmetry in the data,
an unbinned maximum likelihood fit was constructed, using
four inputs for each $\B0 \arr \Jp \Ks$ candidate: 
$\tr$, $\str$, $\QB$ and the 
event-by-event background probability $f_{\rm bac}$, % which is 
derived
from the mass and energy of each candidate (see Section~2).
The total likelihood for an event was given by
\begin{equation} 
{\cal L} = (1-f_{\mathrm{bac}})\cdot {\cal L_{\mathrm{sig}}}
         + f_{\mathrm{bac}} \cdot {\cal L_{\mathrm{bac}}} \; .
\end{equation}
The signal likelihood $\cal{L}_{\mathrm sig}$  was defined as:
\begin{equation} 
{\cal L_{\mathrm{sig}}}(\tr,\str,\QB ; \acp, \dmd, \tau^0) = 
   {\cal F_{\mathrm sig}}(t) \otimes G(t-\tr,\str) \; ,
\end{equation}
where $t$ is the true proper decay time, and $G(t-\tr,\str)$
is a Gaussian representing the proper decay time resolution.
The true proper time distribution is given by
\begin{equation} 
{\cal F_{\mathrm sig}}(t ;\acp, \dmd, \tau^0 ) = \frac{\exp(-t/\tau^0)}{\tau^0} \cdot 
  (1 - \QB \,\acp \sin \dmd t) .
\label{fsig}
\end{equation}
%where the production flavour mistag is included.
The $\B0$ lifetime, $\tau^0$, was taken as $1.56 \pm 0.06$~ps\ \cite{pdg96},
and $\dmd$ was taken as 
$0.467\pm 0.022^{+0.017}_{-0.015}$~ps$^{-1}$ \cite{dilep}.
The likelihood for the
background, ${\cal L_{\mathrm{bac}}}$,
 is defined in the same way, with the true proper time distribution:
\begin{equation} 
{\cal F_{\mathrm{bac}}}(t\ ; \tau_{\mathrm bac}) 
= \frac{\exp(-t/\tau_{\mathrm{bac}})}
{\tau_{\mathrm{bac}}} \; .
\label{fbac}
\end{equation}
The background is dominated by $\bbbar$ events,
and is assumed to have no CP asymmetry.
% although this is addressed
Possible bias due to this assumption is treated
as a systematic error.
%The parameter $\tau_{\mathrm{bac}}$ was taken to be
%2.0~ps.
The effective background lifetime, $\tau_{\mathrm{bac}}$,
was taken to be 2.0\,ps from the Monte Carlo background sample.
This value is larger than
the average b lifetime of 1.55\,ps\,\cite{pdg96}
because the energy of the b hadron is systematically underestimated 
for the background events, 
since the tracks assigned to the $\Jp \Ks$ candidate
do not, in general, include all 
the b-hadron decay products, and include fragmentation products.
%as some tracks from the B decay 
%are not reconstructed.
A large variation of $\pm 0.4$\,ps in 
this parameter is considered in the systematic errors.

Fitting the data for the single parameter \acp,
gave the result
\[ 
\acp = 3.2 _{-2.0}^{+1.8} \; . 
\]
The corresponding $\Delta \log \cal L$ distribution is shown in
Figure~\ref{f:fit}(b). 
It can be parametrised as
\begin{eqnarray*}
-\Delta \log {\cal L} & =  0.116 (\acp-3.2)^2 + 0.00224(\acp-3.2)^4  
  \;\;\; & \acp <3.2 \\
-\Delta \log {\cal L} & =  0.125 (\acp-3.2)^2 + 0.00985(\acp-3.2)^4  
  \;\;\; & \acp >3.2 \; .
\end{eqnarray*}
The parametrisation can be used to combine this result with 
future results from other experiments.
To compare the fitted result with the data,
an estimator, $A$, of the $\B0 \arr \Jp\Ks$ asymmetry
(corrected for the average dilution in each time bin) is shown
in Figure~\ref{f:fit}(c) with the fit result superimposed,
where 
\begin{equation} 
A = \frac{\sum{(1-f_{\mathrm{bac}})\cdot\QB}}
 {\sum{(1-f_{\mathrm{bac}})^2\cdot\QB^2}} \; ,
\end{equation}
and the summations are over all the events in a given time bin.
The large observed values of $A$, typically exceeding the physical
range of the asymmetry, are due to the tagging dilution
factors and, to a lesser extent, the background fraction.
%Large values of asymmetry are caused by 
%corrections for the tagging dilution
%and the background fraction.
%The raw asymmetry, $A_{\mathrm raw}$,
%(not corrected for dilution or background) 
%is shown as a function of time
%in  Figure~\ref{f:fit}(b), where
%\begin{equation} 
%A_{\mathrm raw} = \frac{\sum{(1-f_{\mathrm{bac}})\cdot\QB}}
% {\sum{|(1-f_{\mathrm{bac}})\cdot\QB |}} \; .
%\end{equation}
%
\epostfig{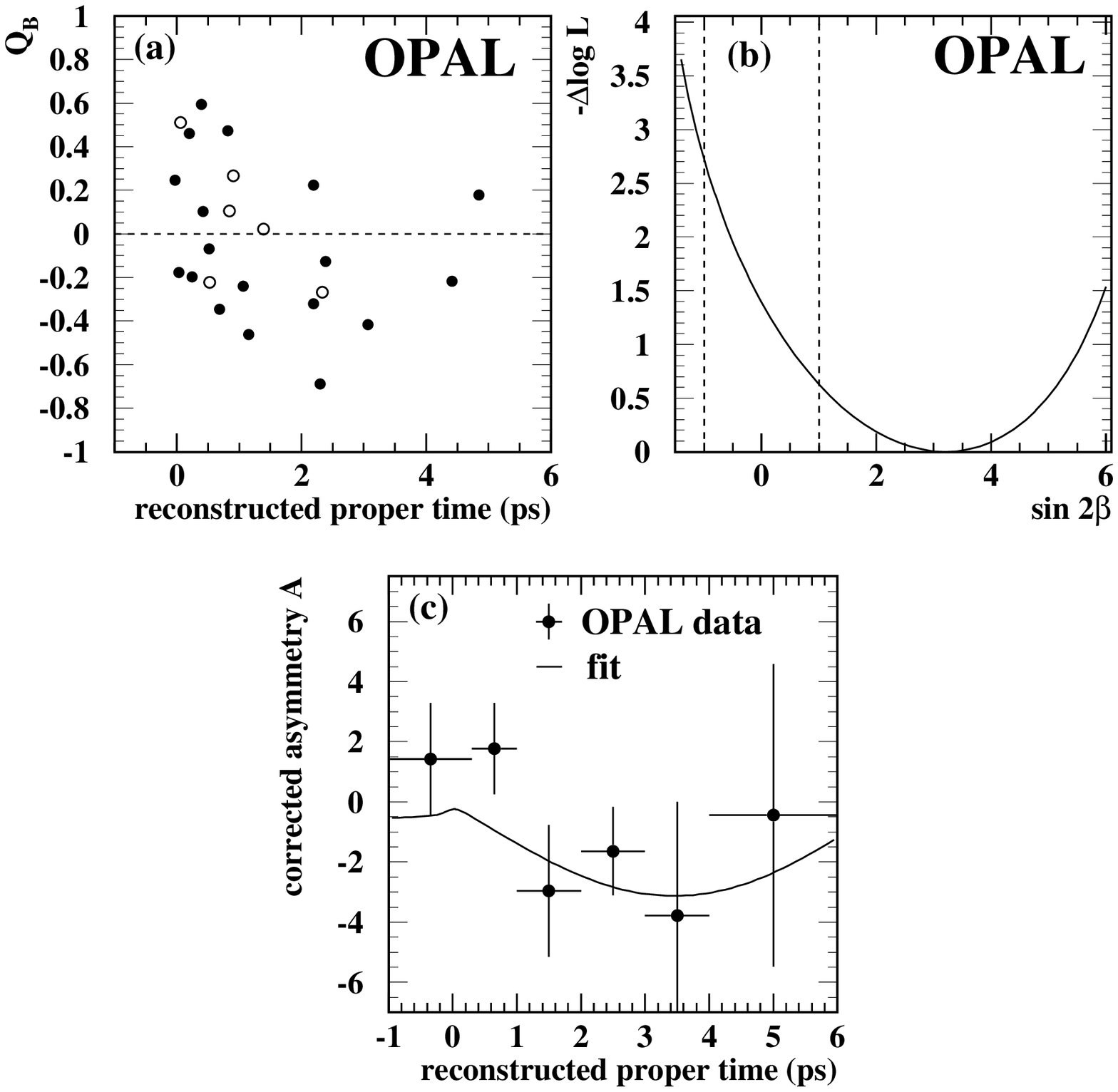}{f:fit}{
(a) The distribution of $\QB$ versus $\tr$ for the $\Jp\Ks$
candidates with invariant masses in the range 5.15--5.40\,\Gcs\ (filled
circles), and 5.00--5.15\,\Gcs\ and 5.40--5.60\,\Gcs\ (open circles);
(b) the $-\Delta\log\cal L$ value as a function of $\sin 2\beta$
from the fit to the 24 data candidates, with the physical
region indicated by the dotted lines;
(c) the distribution of the corrected asymmetry, 
$A$, versus $\tr$
for the $\Jp\Ks$ data, with the fit result superimposed.}
%  
%One can limit $\sin 2\beta$ to physical values by fitting
%the angle $2\beta$ rather than $\sin 2\beta$, giving
%$2\beta = (90 \pm 74)^\circ$.
%The likelihood curve is equivalent to that of the $\sin 2\beta$ fit,
%but with the range limited to $| \sin 2\beta | < 1$.
%Note also that since the angle enters the fit only through its sine,
%there is an ambiguity in quoting this angle.

\section[sys]{Systematic errors and cross checks}
The main sources of systematic error and their effect on the measurement of
$\acp$ are listed in Table~\ref{t:syst}. 
The fit result is sensitive to the level and possible CP asymmetry of
the background, the accuracy of the decay time reconstruction and the 
production flavour tagging dilution. 
%The following sources of systematic error
%have been considered:
\begin{itemize}
\item
The event-by-event purities of the 24 candidates 
have significant statistical errors from the background fit
described in Section~2.
The three fitted parameters
(the number of signal candidates, the
position of the $\B0$ mass peak
and the Peterson parameter for the background energy
distribution)
were each varied by their
statistical errors, one at a time, and the effect
on \acp\ determined.
As the correlations between these parameters were found
to be less than 20\%, these effects were added in quadrature,
%was estimated to be ($52\pm 8$)\%, where
%the error is due to the data statistics. 
%This background uncertainty
leading to a total
error of $^{+0.06}_{-0.07}$ on \acp .

\item
The background mass distribution 
is taken from a fit to a Monte Carlo sample
with four times the statistics of the data sample. Various different
parametrisations and the binned Monte Carlo distribution itself were tried.
As an alternative method,
the shape of the background was taken to be a falling exponential,
with both the normalisation and decay constant of the background being
fitted to the data, as in~\cite{rison}. 
This fit predicts a 
signal purity of 52\% compared to 60\%. 
The Monte Carlo predicts a significant
departure from the exponential shape due to decays of the type
$\mathrm B \arr \Jp \mathrm K^{(*)} X$, followed by 
${\mathrm K^*} \arr \Ks \pi$ when $\mathrm K^*$ are
produced.
Including this contribution explicitly and letting the fitted
exponential describe the remaining background gives a purity of 53\%.
The data show a deficit of events in the region 4.7--4.9\,\Gcs ,
possibly indicating that the Monte Carlo overestimates the background
in this region. 
A further fit to the data alone was therefore
performed, using only the data above 4.9\,\Gcs, with the background
described by a falling exponential,
resulting in a signal purity of 73\%.
The largest variation in \acp\ resulting from these different
parametrisations was found to be $^{+0.25}_{-0.32}$. %and this was taken as
An uncertainty of $\pm 0.32$ was taken for the systematic error due to
the background parametrisation.

\item
Uncertainty on the assumed mass and energy distributions
for the signal also affects
the result of the background fit.
Monte Carlo events with tracking resolution degraded by 10\% were
used to parametrise the signal mass distribution, 
resulting
in a shift in the fitted value of \acp\  of 0.07.
The functional form used to fit the signal mass distribution
was changed from three Gaussian functions to two,
one for the peak of the distribution and another for the
tail.
This caused a shift of 0.11 on \acp .
The uncertainty on the $\B0$ energy distribution was 
assessed by varying the Peterson parameter $\epsilon$ to
cause a change in the mean scaled energy of 0.02, larger than 
the uncertainty on the mean scaled energy of B hadrons \cite{lepxeb}.
The effect on \acp\ was negligible.
 
\item
The final state from the decay
$\B0 \arr \Jp \mathrm K^{*0}$ followed by 
${\mathrm K^{*0}}\arr \Ks \pi^0$ is expected 
to be mainly CP even (i.e.~opposite
CP to the $\Jp \Ks$ final state),
and so could give rise to a possible CP asymmetry in the background.
The contribution from such decays in the signal region was estimated 
from Monte Carlo to be 0.7 events,
and is indicated in Figure~\ref{f:mass}.
If such a contribution had a maximal asymmetry, the effect
on the fitted \acp\ would be 0.03.
The contribution to the background CP asymmetry from
$\B0$ decays involving $\mathrm K^0_L$ mesons was
found to be negligible.
%
%To test the effect of a possible CP asymmetry in the background,
%the contribution from 
%$\B0 \arr \Jp \mathrm K^{*0}$ followed by 
%${\mathrm K^{*0}}\arr \Ks \pi^0$
%in the signal region was estimated to be 0.7 events from Monte Carlo
%(see Figure~\ref{f:mc}).
%The final state is expected to be mainly CP even (i.e. opposite
%CP to the $\Jp \Ks$ final state), so could exhibit an asymmetry.
%If such a contibution had a maximal asymmetry, the effect
%on the fitted \acp\ would be only 0.03.
%The possible contribution from $\mathrm K^0_L$ decays 
%was found to be negligible.

\item
The event-by-event proper time resolution $\str$ is used in the
likelihood fit. Monte Carlo studies indicate that the distribution of
errors in reconstructed proper time divided by $\str$ is
well described by a Gaussian with zero mean and width $1.15\pm 0.15$. 
%A scale factor of 1.15 was therefore applied to the data $\str$ values
%used in the fit, and 
If the proper time resolution is scaled by 1.3, 
the resulting change in \acp\
is $0.01$.

\item
% Scaling. Data has larger widths than MC for qopp,qsame,qvtx by
%4%,1% and 2% respectively. So scale down data qopp,qsame,qvtx by these
%factors as a systematic.
%Systematic on sin2beta is shift of +0.06
The description of $\Qop$ and $\Qvtx$ by the Monte Carlo
was tested
by comparing the correlation of these charges with the lepton charge
in inclusive lepton events in data and Monte Carlo,
as described in Section~3.
The uncertainty on the $\QB$ values was assessed by
scaling the $\Qop$ and $\Qvtx$ values independently
by 1.16 and 1.12, respectively.
These scalings correspond to the sum in quadrature of the
differences seen between data and Monte Carlo and
the statistical precision of the comparisons.
The uncertainty on the modelling of $\Qsam$ was assessed using
$\DSPM \ell$ data as described in Section~3.
The systematic uncertainty was determined by scaling the values
of $\Qsam$ by 1.2, again
corresponding to the quadrature sum of the 
difference seen between data and Monte Carlo and the
statistical precision of the 
comparison.
The total systematic error on \acp\ from these
effects is $_{-0.26}^{+0.31}$.
%by comparing the widths of these distributions
%in inclusive lepton events in data and Monte Carlo.
%The distributions of $\Qop$, $\Qsam$ and $\Qvtx$ were seen to be
%wider in data by 4\%, 1\% and 2\% respectively.
%To evaluate the possible systematic error caused by these
%discrepancies, all $\QB$ values were scaled down by 4\%
%resulting in a change of 0.14 in $\sin 2\beta$.

\item 
%The jet and vertex charges used in the production flavour tagging have
%a bias towards positive values due to the material in the detector and
%asymmetries in the reconstructed track momentum spectra for positive
%and negative tracks. 
%These asymmetries were measured and corrected for
%in the  Monte Carlo when deriving the mistag reference distributions
%(Figure~\ref{f:mist}), and in the data when deriving the mistag $\QB$
%for each event. 
The offsets applied to 
$\Qvtx$ and $\Qop$ were determined from data as
described in Section~3.
These were varied by their
statistical uncertainties.
The offset to the $\Qsam$ jet
charge was determined from Monte Carlo, %as discussed in Section~3.
and checked using the $\B0 \arr \DSPM \ell$ candidates.
The offset was varied by the statistical precision of this test.
%To determine the systematic error, the mean values of $\Qsam$
%were compared for 
%$\DSPM \ell$ samples
%inclusive lepton samples 
%in data and Monte Carlo.
%The difference was treated as a systematic uncertainty, and this
The effect of these variations results in changes in \acp\
%total systematic 
%results in an 
%uncertainty resulting from these variations is
of $_{-0.08}^{+0.14}$.

\item
The performance of the vertex charge algorithm is sensitive to the
tracking resolution. 
The Monte Carlo has been tuned to reproduce the
data impact parameter resolutions as a function of $\cos\theta$, $p$
and the different sub-detectors contributing to a track
measurement. Residual uncertainties were estimated by degrading the
resolution of all tracks by 10\% using a simple smearing technique.
The neural network training and mistag parametrisations were repeated
on this degraded sample, which was then used to derive 
the $\QB$
values that enter the fit for \acp\ . 
The resulting change in \acp\  was $0.01$.

\item
The values for $\dmd$ and the $\B0$ lifetime were varied within their
errors to give the uncertainties listed in the table.
The value of $\tau_{\mathrm{bac}}$ was varied by a conservative
0.4~ps (the difference between the predicted Monte Carlo background
lifetime and the average B meson lifetime)
to allow for uncertainties on the B energy mismeasurement.
\end{itemize}

The total systematic error
is thus $\pm 0.5$. Many of the sources of error have a statistical
component, and many of them scale with the fitted value of
$\sin 2\beta$.
The systematic error
would thus decrease in an analysis with higher statistics.

\begin{table}[tp]
\centering
\renewcommand{\arraystretch}{1.2}
\begin{tabular}{l|c}\hline\hline
Source & $\delta(\acp)$ \\ \hline
Background level (data statistics) & $_{-0.07}^{+0.06}$ \\
Background shape & $\pm 0.32$ \\
Signal shape & $\pm 0.13$ \\
Background asymmetry & $\pm 0.03$ \\
Proper time reconstruction & $\pm 0.01$ \\
Jet and vertex charge modelling & $_{-0.26}^{+0.31}$ \\
\vspace{1mm}
Jet charge offsets & $^{+0.14}_{-0.08}$ \\
Vertex charge performance & $\pm 0.01$ \\ 
$\dmd$ value  &  $\pm 0.10$ \\
$\B0$ lifetime  &  $\pm 0.01$ \\
Background lifetime ($\pm 0.4$~ps)  & $_{-0.02}^{+0.01}$ \\ \hline
Total & $_{-0.46}^{+0.50}$ \\
\hline
\end{tabular}
\caption{\label{t:syst} Sources of systematic error in the measurement
  of \acp .}
\renewcommand{\arraystretch}{1.}
\end{table}

A number of consistency checks were also performed. The result was
found to be stable when the least well tagged events 
(those with $|\QB|<0.25$), the events with highest background
($f_{\mathrm bac}>0.5$), or the events outside the purest mass
region (5.15--5.40\,\Gcs ) were removed.
The values of $\sin 2\beta$ resulting from these checks 
were found to be $4.0^{+1.9}_{-2.3}$\ ,
$3.2^{+1.8}_{-2.0}$ and $2.8^{+1.9}_{-2.0}$ respectively.
The data were fitted for the $\B0$ lifetime,
giving a result of $1.2 ^{+0.5}_{-0.4}$~ps
(independent of $\sin 2\beta$),
consistent with the world average.
In addition, the assumption that the background
exhibits no CP asymmetry was tested by repeating the fit, using only
events in the sideband region and setting $f_{\mathrm{bac}}$
to 0 for every event.
The fitted value of $\sin 2\beta$ in this case was 
$0.38 ^{+0.45}_{-0.49}$, consistent with zero. The
selection cuts were loosened to give a sideband data sample three times
larger, and a fitted value of $\sin 2\beta$ of $-0.12\pm 0.30$ was obtained.
The background and CP asymmetry fits were also repeated on a 
Monte Carlo sample with no CP violation and four times the data
statistics,  giving asymmetries consistent with zero for both signal
and sideband regions.
The $\B0\arr\DSPM\ell$ sample was used to perform a further cross
check for the absence of large biases in the determination of $\QB$.
The events were separated according to the sign of $\Ql$, and the
average $\QB$, $\langle \QB^+ \rangle$ and 
$\langle \QB^- \rangle$, was calculated for each
subsample. 
To account for charge biases due to the $\DSPM$ selection,
the average value of $\QB$ was calculated as 
$(\langle \QB^+ \rangle + \langle \QB^- \rangle )/2$
and was found to be $-0.016\pm 0.011$, consistent with zero.
The Monte Carlo prediction for the
$\Qop$ offset, which is not used in the analysis,
disagrees with the value determined from the data.
If the fit is repeated taking all offsets
from the Monte Carlo, $\sin 2\beta$ is shifted
by $-0.28$.
This discrepancy between data and Monte Carlo 
does not affect the description of the tagging dilution.

% MC background fit
%The technique of the background fit was tested on a Monte
%Carlo sample roughly four times larger than the data sample.
%Using a simple exponential for the background mass distribution,
%the fitted purity was $(52 \pm 5)\%$ compared to a true purity
%of 60\%.
%If the shape of the background was modified to include
%peaking contributions, as described above in the systematic errors,
%the fitted purity was $(55 \pm 5)\%$.
%The fitted value of $\sin 2\beta$ was consistent with
%zero in this sample.

The value of $\sin 2\beta$ can also be estimated from 
the time-integrated asymmetry.
In this case, the lower limit of the time integration can be varied to
optimise the sensitivity --- i.e.~the ability to distinguish different 
true values of $\sin 2\beta$.
For data samples of this size and purity, the optimum lower
bound\footnote{
The value is smaller
than that which would be obtained in the absence of background.}
was found,
using Monte Carlo studies,
to be 0.7\,ps.
The value of $\sin 2\beta$ obtained from our
data sample using this method is $2.0^{+1.1}_{-1.5}$,
where 12 events are included in the range of integration. 
The probability
of obtaining time dependent and time integrated measurements
disagreeing at this level or more was found to be 20\%.
Monte Carlo studies indicate that the errors obtained from both
types of fit increase as the central values deviate from zero, and that
the time dependent fit yields smaller errors on average. They also show that
the time dependent fit has a greater sensitivity to the true value of
$\sin 2\beta$ than the time integrated method, even after optimising the
lower time-integration bound.

\section[con]{Discussion and conclusion}\label{s:disc}

The result from this analysis can be 
interpreted by calculating the probabilities 
to see a deviation, in the positive $\sin 2\beta$ direction,
at least as far from the true value as that observed, 
for different true values of $\sin 2\beta$. 
The deviation is defined by the difference in $\log \cal{L}$ between
the fitted value and the assumed true value. 
This definition is used because the sensitivity varies from experiment
to experiment.
Monte Carlo samples of 24 
candidates, with the same background and tagging distributions as those
expected in the data, were generated to determine these probabilities.
The probabilities for the $\log \cal{L}$ differences seen in the
data, with correction for the systematic error, were found to be 
be 1.6\%, 7.8\% and 21.3\% if the true value of $\sin 2\beta$
were $-1$, 0 and +1, respectively. 
These probabilities indicate the consistency of the result with these
values of $\sin 2\beta$, and should not be interpreted as confidence levels.
The distributions of fitted 
$\sin 2\beta$ for Monte Carlo experiments with true $\sin 2\beta$ 
of $-1$, 0
and +1 are shown in Figures~\ref{f:ltoy}(a), (b) and (c).

\epostfig{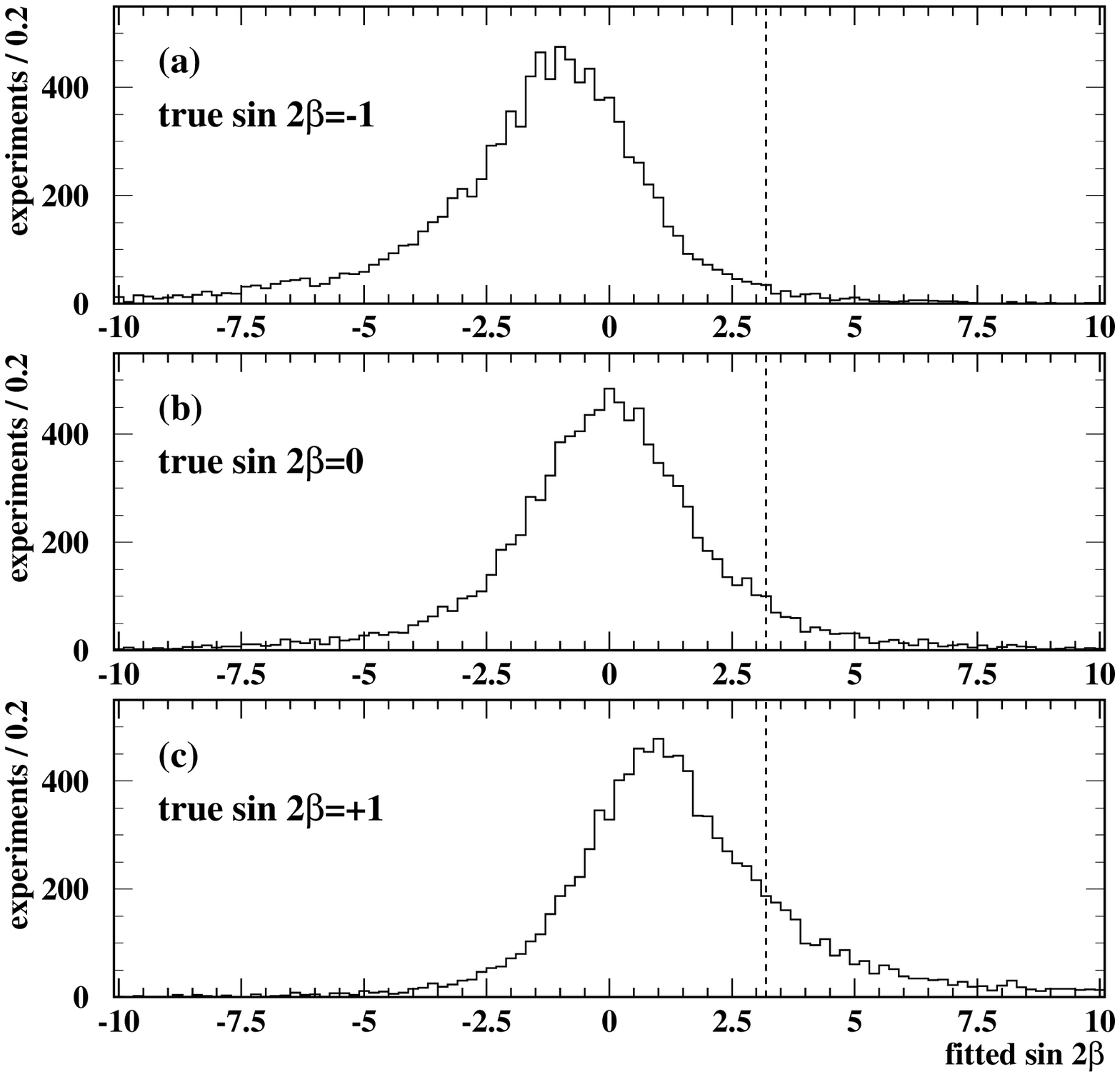}{f:ltoy}{
The distributions of
fitted $\sin 2\beta$ in Monte Carlo samples of 24 events. 
The dotted
lines indicate the data fitted value.
}

An alternative interpretation is given by the Bayesian approach \cite{bayes}
assuming equal {\em a priori\/} probabilities for every possible true value of 
$\sin 2\beta$. In this case, the probabilities for $\sin 2\beta$ to 
be greater or less than zero, with correction for the systematic
error, are found to be 68.5\% and 31.5\% respectively.

In conclusion, 
the time dependent CP asymmetry in the decays $\B0\arr\Jp\Ks$ and 
$\Bobar\arr\Jp\Ks$ has been measured using data collected with
the OPAL detector at LEP between 1990 and 1995. 
From 24 reconstructed
$\B0\arr\Jp\Ks$ candidates with a purity of about 60\%,
the CP violation amplitude, which
is \acp\ in the Standard Model,
has been found to be:
\[
\acp=3.2 ^{+1.8}_{-2.0}\ \pm 0.5 \; ,
\]
where the first error is statistical and the second systematic. 
The systematic error has a large statistical component, and much
of it scales with the central value.

This is the first direct study of the 
CP asymmetry in the $\B0\arr\Jp\Ks$ system.
It can be combined with
other results in the future by using the log-likelihood curve given
in section~4.

\par
\vspace*{1.cm}
\section*{Acknowledgements}
%\noindent{\bf\Large Acknowledgements}
%\par
\noindent
We particularly wish to thank the SL Division for the efficient operation
of the LEP accelerator at all energies
 and for
their continuing close cooperation with
our experimental group.  We thank our colleagues from CEA, DAPNIA/SPP,
CE-Saclay for their efforts over the years on the time-of-flight and trigger
systems which we continue to use.  In addition to the support staff at our own
institutions we are pleased to acknowledge the  \\
Department of Energy, USA, \\
National Science Foundation, USA, \\
Particle Physics and Astronomy Research Council, UK, \\
Natural Sciences and Engineering Research Council, Canada, \\
Israel Science Foundation, administered by the Israel
Academy of Science and Humanities, \\
Minerva Gesellschaft, \\
Benoziyo Center for High Energy Physics,\\
Japanese Ministry of Education, Science and Culture (the
Monbusho) and a grant under the Monbusho International
Science Research Program,\\
German Israeli Bi-national Science Foundation (GIF), \\
Bundesministerium f\"ur Bildung, Wissenschaft,
Forschung und Technologie, Germany, \\
National Research Council of Canada, \\
Research Corporation, USA,\\
Hungarian Foundation for Scientific Research, OTKA T-016660, 
T023793 and OTKA F-023259.\\

\end{document}